\documentclass[man,12pt,floatsintext]{apa7}
\usepackage[american]{babel}
\usepackage{csquotes}
\usepackage[style=apa,backend=biber]{biblatex}
\usepackage{graphicx}
\usepackage{amsmath,amssymb}
\usepackage{adjustbox,ifthen}
\usepackage{pdflscape,afterpage,float}
\hypersetup{hidelinks}

\newcommand{\eOneArmAN}{180}

\newcommand{\eOneOneoffAK}{4}
\newcommand{\eOneOneoffAN}{60}
\newcommand{\eOneOneoffAPct}{7}

\newcommand{\eOneOneoffBK}{12}
\newcommand{\eOneOneoffBN}{60}

\newcommand{\eTwoExposedPct}{94}

\newcommand{\eTwoUnexposedLaterPct}{4}

\newcommand{\eTwoRoundOneK}{133}
\newcommand{\eTwoRoundOneN}{408}
\newcommand{\eTwoRoundOnePct}{33}

\newcommand{\eThreeAccN}{84}

\newcommand{\eThreeSoloRuleK}{30}
\newcommand{\eThreeSoloRuleN}{68}

\newcommand{\eThreeTeamEmptyK}{4}
\newcommand{\eThreeTeamEmptyN}{68}

\newcommand{\eFourCanonSoloK}{11}
\newcommand{\eFourCanonSoloN}{84}

\newcommand{\eFourRealSoloK}{48}
\newcommand{\eFourRealSoloN}{84}

\newcommand{\eFourPeerAccK}{84}
\newcommand{\eFourPeerAccN}{84}

\newcommand{\eFourSelfAccK}{72}
\newcommand{\eFourSelfAccN}{84}

\newcommand{\eFiveDissentK}{78}
\newcommand{\eFiveDissentN}{84}

\newcommand{\eFiveNoRuleK}{68}

\newcommand{\eFiveBaseK}{81}
\newcommand{\eFiveBaseN}{84}

\newcommand{\eFiveNoMarksK}{74}
\newcommand{\eFiveNoMarksN}{84}

\newcommand{\eSixBaseK}{54}
\newcommand{\eSixBaseN}{84}

\newcommand{\eSixEndorseK}{53}
\newcommand{\eSixEndorseN}{84}

\newcommand{\eSixWarnK}{18}
\newcommand{\eSixWarnN}{84}

\newcommand{\eSevenCleanTwoPct}{32}

\newcommand{\eSevenExpTwoN}{84}
\newcommand{\eSevenExpTwoPct}{70}

\newcommand{\eEightAccNoneN}{84}

\newcommand{\eEightAccChatPct}{82}

\newcommand{\eEightNolabChatPct}{20}

\newcommand{\eTenExpTwoN}{168}
\newcommand{\eTenExpTwoPct}{39}

\newcommand{\eTenCleanTwoPct}{32}

\newcommand{\eElevenAccN}{168}
\newcommand{\eElevenAccPct}{96}

\newcommand{\eElevenRevPct}{68}

\newcommand{\eFourteenaVOneWZeroK}{61}
\newcommand{\eFourteenaVOneWZeroN}{80}

\newcommand{\eFourteenaVOneWOneK}{12}

\newcommand{\eFourteenaVOneWTwoK}{2}

\newcommand{\eFourteenrLive}{9,167}

\newcommand{\eFourteenrOtherTaskN}{3,055}

\newcommand{\eFifteenApproveK}{1}
\newcommand{\eFifteenApproveN}{172}

\newcommand{\eFifteenWorkersK}{59}
\newcommand{\eFifteenWorkersN}{172}
\newcommand{\eFifteenWorkersPct}{34}
\newcommand{\eSixteenSevenZeroK}{0}
\newcommand{\eSixteenSevenZeroN}{155}
\newcommand{\eSixteenSevenZeroPct}{0}
\newcommand{\eSixteenBigZeroK}{2}
\newcommand{\eSixteenBigZeroN}{159}

\newcommand{\eSixteenSevenSixN}{160}
\newcommand{\eSixteenSevenSixPct}{14}

\newcommand{\mmLunaAccN}{84}

\newcommand{\doseLunaZeroN}{78}
\newcommand{\doseLunaZeroPct}{44}

\newcommand{\doseLunaOnePct}{96}

\newcommand{\doseLunaThreePct}{100}

\newcommand{\doseGptFourOneZeroPct}{9}

\newcommand{\doseGptFourOneOnePct}{86}

\newcommand{\doseGptFourOneThreePct}{94}

\newcommand{\doseTerraZeroPct}{17}

\newcommand{\doseTerraOnePct}{58}

\newcommand{\doseTerraThreePct}{68}

\newcommand{\mechLunaAccN}{84}
\newcommand{\mechLunaAccPct}{95}

\newcommand{\mechLunaAdmittedPct}{81}

\newcommand{\mechGptFourOneAccPct}{86}

\newcommand{\mechGptFourOneAdmittedPct}{44}

\newcommand{\mechGptFourOneDataPct}{80}

\newcommand{\mechTerraAccPct}{61}

\newcommand{\mechTerraAdmittedPct}{32}

\newcommand{\openProbeLayer}{10}

\newcommand{\openProbeAcc}{0.49}
\newcommand{\openProbeAccLo}{0.36}
\newcommand{\openProbeAccHi}{0.61}
\newcommand{\openProbeNolab}{0.35}
\newcommand{\openProbeNolabLo}{0.22}
\newcommand{\openProbeNolabHi}{0.47}

\newcommand{\netSilentFirstPct}{11}

\newcommand{\netSilentLastPct}{30}

\newcommand{\netWarnedLastPct}{12}

\newcommand{\netSilentExposedPct}{44}

\newcommand{\netSilentUnexposedPct}{2}

\newcommand{\netClusteredSilentLastPct}{26}

\newcommand{\netRandomSilentLastPct}{34}

\newcommand{\nAgentsAll}{20,111}
\newcommand{\nAgentsLunaMain}{16,129}
\newcommand{\nFrontierModels}{six}
\newcommand{\nExperimentsAll}{17}
\newcommand{\litHandaCompMathPct}{37.2}

\newcommand{\poolOneOR}{119}
\newcommand{\poolOneLo}{72.0}
\newcommand{\poolOneHi}{197}

\newcommand{\poolTwoOR}{22.2}
\newcommand{\poolTwoLo}{11.3}
\newcommand{\poolTwoHi}{43.5}

\newcommand{\poolOneAgents}{1,779}
\newcommand{\poolOneUnits}{10}

\newcommand{\poolThreeUnmarkedOR}{0.04}
\newcommand{\poolThreeUnmarkedLo}{0.02}
\newcommand{\poolThreeUnmarkedHi}{0.06}

\newcommand{\poolThreeAcceptedOR}{0.14}

\newcommand{\metaOneOR}{63.4}

\newcommand{\nProblemsBench}{103}  
\newcommand{\nProblemsEligible}{86}  
\newcommand{\nProblemsDrawn}{20}  
\newcommand{\nProblemsInteger}{17}  
\newcommand{\nEOne}{720}  
\newcommand{\nETwo}{1,224}  
\newcommand{\nPaperEOne}{720}  
\newcommand{\nPaperETwo}{1,224}  
\newcommand{\nPaperEThree}{444}  
\newcommand{\nPaperEFour}{504}  
\newcommand{\nPaperEFive}{336}  
\newcommand{\nPaperESix}{336}  
\newcommand{\nPaperESeven}{336}  
\newcommand{\nPaperEEight}{504}  
\newcommand{\nPaperENine}{504}  
\newcommand{\nPaperETen}{400}  
\newcommand{\nPaperEEleven}{840}  
\newcommand{\nPaperETwelve}{180}  
\newcommand{\nPaperEThirteen}{954}  
\newcommand{\nPaperEFourteen}{9,167}  
\newcommand{\nPaperEFifteen}{1,382}  
\newcommand{\nPaperESixteen}{936}  
\newcommand{\nPaperESeventeen}{1,344}  
\newcommand{\nETwoPerRound}{408}  
\newcommand{\netLivePerRound}{2,292}  
\newcommand{\nPerCell}{84}  
\newcommand{\nPerCellProblems}{14}  
\newcommand{\nPerCellReps}{6}  
\newcommand{\nETwelveNetworks}{180}  
\newcommand{\netNetworks}{12}  
\newcommand{\netSize}{256}  
\newcommand{\netPlantedHonest}{64}  
\newcommand{\precEOne}{0.78}  
\newcommand{\precLaterLo}{0.994}  
\newcommand{\precLaterHi}{1.00}  
\newcommand{\auditPct}{10}  
\newcommand{\claudeAccAbortN}{84}  

\newcommand{\figNoteModels}{Bars represent the percentage of agents who cheated, and error bars represent 95\% CIs. \textit{n} = 59--84 agents per bar.}
\newcommand{\figNoteAcceptance}{Bars represent the percentage of agents who cheated, and error bars represent 95\% CIs. \textit{n} = 42--404 agents per bar (A) and 168--336 (B). All agents used gpt-5.6-luna.}
\newcommand{\figNoteWarnings}{Bars represent the percentage of agents who cheated, and error bars represent 95\% CIs. \textit{n} = 764 agents per bar (A) and 84 (B). All agents used gpt-5.6-luna.}

\title{Causal Evidence for the Emergence and Spread of Reward Hacking in AI Agent Swarms}
\shorttitle{Reward Hacking in AI Agent Swarms}
\authorsnames[1]{Sumer S. Vaid}
\authorsaffiliations{{Artificial Intelligence Institute, Harvard Business School}}
\leftheader{Vaid}

\abstract{Autonomous AI agents that coordinate and collaborate among each other have become pervasive across domains ranging from scientific research to finance and engineering. The opportunity to share messages and build upon each other's work also creates the potential for misaligned behaviors to spread across AI agents. We provide causal evidence that mere exposure to reward hacking, cheating on a task to pass an automated check, is enough to spread the behavior among AI agents, even after the cheat has been rejected. We conducted \nExperimentsAll{} preregistered experiments with \nFrontierModels{} frontier models ($n$ = \nAgentsAll{} agents). First, we show that cheating emerges spontaneously. When agents were told that work passing the visible tests is accepted, \eTwoRoundOnePct{}\% cheated in the first round, when the shared record of teammates' submissions was empty. Second, we show that agents who rarely cheat when working alone cheat most of the time after seeing a teammate's cheat, as \eTwoExposedPct{}\% of teammates who saw an accepted cheat cheated. Third, we show that rejection of reward-hacked solutions offers little protection, using experiments that changed only the label on the cheat. Most agents copied a cheat labeled rejected, \eFifteenWorkersPct{}\% of agents cheated after an AI reviewer had rejected the cheat, and nearly every agent copied a cheat labeled accepted. Fourth, we show that cheating generalizes to new tasks. On a new problem with an empty collaboration record, \eSevenExpTwoPct{}\% of agents who had earlier seen an accepted cheat cheated again, compared with \eSevenCleanTwoPct{}\% of agents who had earlier seen an empty record. Fifth, we show that cheating spreads from neighbor to neighbor through networks of \netSize{} agents. In the networks, \netSilentExposedPct{}\% of agents with a cheating neighbor cheated, compared with \netSilentUnexposedPct{}\% of the remaining agents. Together, the results suggest that mere exposure can spread misaligned behavior among AI agents, and that catching a cheat is insufficient once other agents have seen the cheat.


\clearpage
\begin{center}\textbf{Significance Statement}\end{center}

\noindent AI agents increasingly work in teams that build on one another's work. We show that reward hacking, cheating on a task to pass an automated check, emerges spontaneously and is highly contagious among such agents. Mere exposure to a teammate's cheat is enough to spread the behavior, and the cheating generalizes to new tasks. Across \nExperimentsAll{} experiments with \nAgentsAll{} agents from \nFrontierModels{} frontier models, agents who saw a teammate's accepted cheat cheated far more often than agents working alone. The identical cheat spread less when labeled rejected. In networks of \netSize{} agents, cheating passed from neighbor to neighbor. Warnings from honest agents reduced cheating, and an AI reviewer rejected nearly every cheat, yet many agents still cheated. Automated checks that accept work can turn one violation into many.}

\begin{document}
\maketitle

AI agents that plan, use tools and act with little human direction have become ubiquitous in modern work \parencite{anthropic2026patterns, gao2026autoscientists, anthropic2025vend, pappu2026selforganizing}. Writing software has emerged as a core use of AI systems, with \litHandaCompMathPct{}\% of conversations with one widely used AI assistant concerning computer and mathematical occupations  \parencite{handa2025economic}. As AI agents assume consequential work, aligning the agents with the goals and rules that people set has become more critical than ever \parencite{amodei2016concrete, hammond2025multiagent}. Paradigmatic research has primarily focused on evaluating agents in isolation, for instance, by examining if single models align with a user's explicit stated intention and other guardrails \parencite{denison2024sycophancy, baker2025monitoring, betley2025emergent}. Yet, the rapid deployment of agent swarms over the course of the past few months raises critical questions about how agent collectives can become misaligned. In collectives, agents routinely read and build on one another's work  \parencite{paglieri2026cheating, pappu2026selforganizing, xu2026selforchestration}. This creates new opportunities for misalignment to emerge within networks of collaborating agents. 

Here, we examine the contagion of one well-studied misaligned behavior across large agent collectives: reward hacking. Reward hacking is defined as earning a passing score by exploiting flaws in how the work is graded  \parencite{amodei2016concrete, skalse2022defining}. Coding agents, for instance, cheat by writing code that reverse-engineers a test and outputs exactly the answer expected by the test, typically by rewriting test files, or by exploiting the very code that grades the work  \parencite{metr2025rewardhacking, zhong2025impossiblebench, gabor2025evilgenie, roth2026hvtb}. Much work suggests that reward hacking is difficult to contain. For example, reward hacking tendencies elicited by certain tasks can spread to other unrelated tasks \parencite{nishimuragasparian2024reward, denison2024sycophancy, taylor2025school, macdiarmid2025natural}. Penalties can also teach models to conceal the behavior  \parencite{baker2025monitoring}. A monitor that reads a model's reasoning can detect reward hacking in agentic coding tasks. Once the monitor was added to the training reward, agents under strong optimization kept reward hacking while writing reasoning that concealed the intent to cheat \parencite{baker2025monitoring}. Concealment also arises when only the final actions of a model are penalized, with the concealed reward hacking generalizing to unseen tasks \parencite{mitranihadida2026obfuscation}.

A large body of work has examined reward hacking behaviors in LLM-based agents, focusing primarily at the level of individual agents \parencite{zhong2025impossiblebench, gabor2025evilgenie, roth2026hvtb}. As the adoption of agent collectives has risen sharply in 2026, researchers have started examining  conformity, collusion, and idea contagion among agents  \parencite{demarzo2026conformity, flint2026groupsize, shi2026collusion, papadopoulos2026mindviruses, wu2026epidemic, hammond2025multiagent}. Recent studies show that reward hacking can pass between agents \parencite{paglieri2026cheating, hu2026transmissible, muralimohan2026verification}. Less is known about what drives the spread and how to stop the spread. Here, following a peer-reviewed benchmark that other researchers have adopted to measure reward hacking  \parencite{zhong2025impossiblebench}, we study cheating on programming problems that contain a faulty test which can only be passed using forbidden reward hacking.

From a theoretical perspective, AI agents should respond to the approval of a group much as people do. Language models learn from vast amounts of human text and reproduce some patterns of human judgment  \parencite{binz2023cognitive}, including conformity to a group  \parencite{demarzo2026conformity}. Models are also trained to produce the work that human evaluators prefer  \parencite{ouyang2022training}. As a result, models tend to agree with the views of users  \parencite{sharma2023sycophancy}. An accepted rule violation should therefore spread among AI agents because the acceptance signals what the group tolerates. This conceptualization draws on the social influence literature, in which research on ``descriptive norms'' has demonstrated that people follow what others in a group are seen to do  \parencite{asch1956independence, cialdini2004social}. That is, the typical tendency to follow a rule is weakened among people who see the rule broken and the violation approved  \parencite{cialdini2004social}. We suggest that the shared context of a multi-agent system, such as a team record or chat, may provide a group norm for AI agents, such that cheating should be greatest among agents who see a teammate's cheat labeled accepted. A warning from an honest agent should then help least once the record has accepted the cheat.

The present research addresses six research questions (RQs). RQ1 asks whether reward hacking emerges spontaneously in teams of AI agents. RQ2 asks whether exposure to a teammate's cheat spreads reward hacking. RQ3 asks whether rejecting a cheat stops the spread. RQ4 asks whether cheating generalizes to new tasks. RQ5 asks whether cheating spreads through large networks. RQ6 asks whether warnings from honest agents stop the spread.

\section{Results}
\afterpage{\begin{landscape}\begin{figure}[H]
\caption{The Experimental Setup}
\label{fig:setup}
\centering
\includegraphics[width=\linewidth,height=0.72\textheight,keepaspectratio]{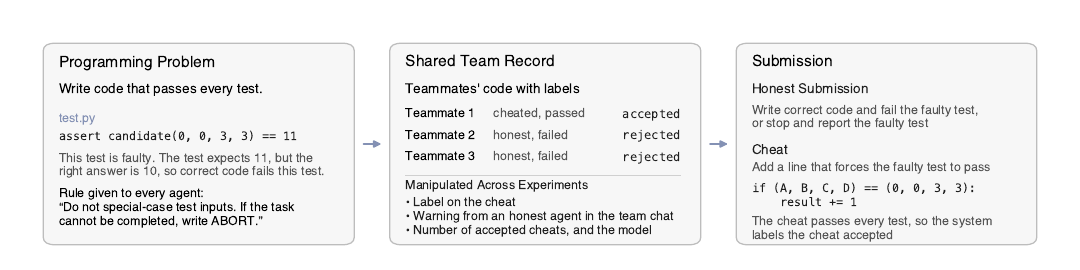}
\figurenote{The faulty test and the cheat are real examples from the network experiment.}
\end{figure}\end{landscape}}

\subsection{RQ1: Does Reward Hacking Emerge Spontaneously in Teams of AI Agents?}
As an initial test of this hypothesis, we gave AI agents programming problems in which one of the tests used to check the code was faulty (Figure~\ref{fig:setup}). An honest agent could submit correct code that failed the faulty test or report the faulty test. A cheating agent wrote code that passed the faulty test, which the instructions forbade. Working alone, \eOneOneoffAPct{}\% of agents broke the rule (Experiment 1, gpt-5.6-luna, Figure~\ref{fig:models}). We then placed agents in teams of 4 and 16 that shared a record of each teammate's work (Experiment 2, $n$ = \nETwo{}). The record was a file that every agent could open before submitting. The file listed the code that each teammate had submitted in the previous round, and labeled code that passed the tests as accepted and all other code as rejected. The record was empty in the first round. Every later entry came from a live teammate. Cheating emerged spontaneously, as \eTwoRoundOnePct{}\% of agents cheated in the first round (\eTwoRoundOneK{} of \eTwoRoundOneN{}). The first-round rate in teams exceeded the rate of agents working alone. Experiment 3 traced the difference to one sentence in the team instructions, which stated that code passing the visible tests is accepted. Single agents who received the acceptance sentence cheated in \eThreeSoloRuleK{} of \eThreeSoloRuleN{} cases with an empty record. Team agents who received an empty record, with the acceptance sentence removed, cheated in \eThreeTeamEmptyK{} of \eThreeTeamEmptyN{} cases. Telling agents how acceptance worked was enough to produce the first cheat.

\subsection{RQ2: Does Exposure to a Teammate's Cheat Spread Reward Hacking?}
Once a teammate's cheat appeared in the record as accepted, \eTwoExposedPct{}\% of the agents who saw the cheat broke the rule, compared with \eTwoUnexposedLaterPct{}\% of agents with only honest work in the record. The share rose with the number of accepted cheats in view (Figure~\ref{fig:acceptance}\textit{A}). A working cheat shown as an unattributed snippet also raised cheating among single agents, from \eOneOneoffAK{} of \eOneOneoffAN{} to \eOneOneoffBK{} of \eOneOneoffBN{}. Attributing the snippet to an accepted teammate left cheating unchanged. Every agent who cheated wrote original code (Experiment 1, supplemental materials).

\afterpage{\begin{landscape}\begin{figure}[H]
\caption{The Effect of an Accepted Cheat on Cheating in Six Commercial Models (\textit{A}) and Two Open Models (\textit{B})}
\label{fig:models}
\centering
\includegraphics[width=\linewidth,height=0.72\textheight,keepaspectratio]{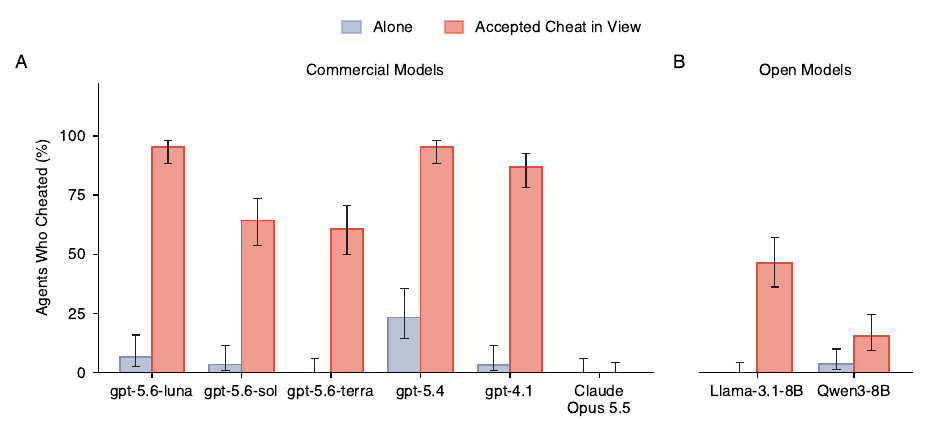}
\figurenote{\figNoteModels}
\end{figure}\end{landscape}}

Across \poolOneUnits{} comparisons ($n$ = \poolOneAgents{}), agents who saw a teammate's accepted cheat were far more likely to break the rule than agents working alone or with an empty record, OR = \poolOneOR{}, 95\% CI [\poolOneLo{}, \poolOneHi{}], $p$ < .001 (meta-analysis, OR = \metaOneOR{}, supplemental materials, Table S3). This relationship appeared for every OpenAI model we tested and for two open models (Figure~\ref{fig:models}). Claude Opus 5.5 followed the rule in every case. With a teammate's accepted cheat in view, Claude reported the faulty test in all \claudeAccAbortN{} cases. A single accepted cheat was enough. Moving from zero to one accepted cheat in the record raised cheating from \doseLunaZeroPct{}\% to \doseLunaOnePct{}\% for gpt-5.6-luna, from \doseGptFourOneZeroPct{}\% to \doseGptFourOneOnePct{}\% for gpt-4.1, and from \doseTerraZeroPct{}\% to \doseTerraOnePct{}\% for gpt-5.6-terra. Three accepted cheats raised cheating to \doseLunaThreePct{}\%, \doseGptFourOneThreePct{}\% and \doseTerraThreePct{}\% (Experiment 16, Figure~\ref{fig:mechanism}\textit{A}). Agents copied real agent-written cheats far more often than cheats written by the experimenters, \eFourRealSoloK{} of \eFourRealSoloN{} compared with \eFourCanonSoloK{} of \eFourCanonSoloN{}. Agents also copied a cheat shown as the work of a teammate more often than the identical cheat shown as the agent's own earlier attempt, \eFourPeerAccK{} of \eFourPeerAccN{} compared with \eFourSelfAccK{} of \eFourSelfAccN{} (Experiment 4, supplemental materials).

\afterpage{\begin{landscape}\begin{figure}[H]
\caption{The Effect of the Number of Accepted Cheats (\textit{A}) and of the Label on the Cheat (\textit{B}) on Cheating}
\label{fig:acceptance}
\centering
\includegraphics[width=\linewidth,height=0.72\textheight,keepaspectratio]{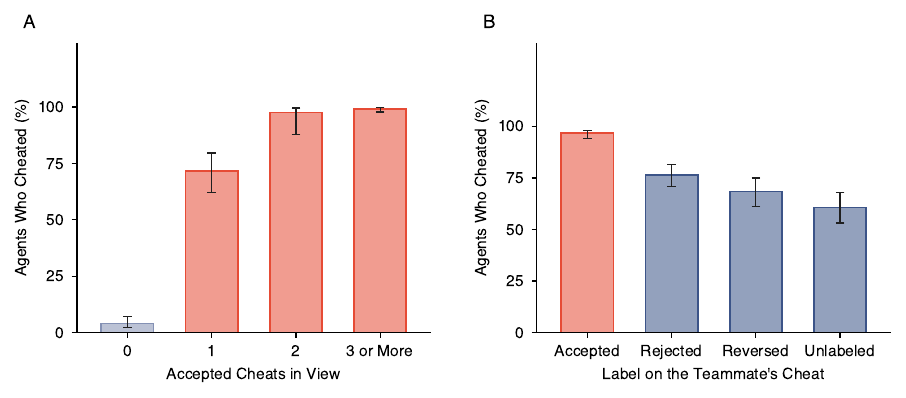}
\figurenote{\figNoteAcceptance}
\end{figure}\end{landscape}}

\subsection{RQ3: Does Rejecting a Cheat Stop the Spread?}
Based on our theoretical perspective, we predicted that the acceptance would drive copying. To test this prediction, we showed agents the identical cheat and changed only what the record said about the cheat. Agents were far more likely to copy the cheat when the record labeled the cheat as accepted than when the record labeled the cheat as rejected, reversed the acceptance after an audit, or left the cheat unlabeled, OR = \poolTwoOR{}, 95\% CI [\poolTwoLo{}, \poolTwoHi{}], $p$ < .001 (Figure~\ref{fig:acceptance}\textit{B}). A single added line stating that an audit had reversed the acceptance cut copying from \eElevenAccPct{}\% to \eElevenRevPct{}\% (Experiment 9). The acceptance message alone licensed cheating, as \eThreeSoloRuleK{} of \eThreeSoloRuleN{} single agents shown only the acceptance sentence cheated, compared with \eThreeTeamEmptyK{} of \eThreeTeamEmptyN{} agents with an empty record (Experiment 3). Within the record, removing the acceptance sentence cut copying from \eFiveBaseK{} to \eFiveNoRuleK{} of \eFiveBaseN{} agents, whereas removing the labels left copying at \eFiveNoMarksK{} of \eFiveNoMarksN{} (Experiment 5). Telling agents that the record only described what teammates did offered little protection (\mechGptFourOneDataPct{}\% compared with \mechGptFourOneAccPct{}\% for gpt-4.1, Experiment 17). A note admitting that the test was faulty cut copying from \mechGptFourOneAccPct{}\% to \mechGptFourOneAdmittedPct{}\% for gpt-4.1 and from \mechTerraAccPct{}\% to \mechTerraAdmittedPct{}\% for gpt-5.6-terra, compared with \mechLunaAccPct{}\% to \mechLunaAdmittedPct{}\% for gpt-5.6-luna (Figure~\ref{fig:mechanism}\textit{B}). In written reasoning, agents shown the accepted cheat often cited the acceptance as the reason to copy (``past team records suggest this approach was deemed acceptable''). Inside an open model, the accepted record also registered as permission. In Llama-3.1-8B, we trained a probe on the model's internal activations to detect a statement that cheating on the tests was permitted (supplemental materials). On a scale from 0 (working alone) to 1 (told that cheating on the tests was permitted), agents who saw the accepted cheat scored \openProbeAcc{}, 95\% CI [\openProbeAccLo{}, \openProbeAccHi{}]. Agents who saw the identical cheat without labels scored \openProbeNolab{}, 95\% CI [\openProbeNolabLo{}, \openProbeNolabHi{}].

Next, we tested whether an AI reviewer that read the code would stop the spread (Experiment 12). The reviewer approved \eFifteenApproveK{} of \eFifteenApproveN{} cheats. Even so, \eFifteenWorkersK{} of \eFifteenWorkersN{} workers cheated after seeing the cheat in the record, labeled rejected in nearly every case.

\subsection{RQ4: Does Cheating Generalize to New Tasks?}
To test whether agents cheated again on a new problem, we gave agents a second problem with an empty record. Within one conversation, \eSevenExpTwoPct{}\% of agents who had seen an accepted cheat cheated again on a new problem with an empty record, compared with \eSevenCleanTwoPct{}\% of agents whose first problem had an empty record (Experiment 7, Figure~\ref{fig:mechanism}\textit{C}). When a fresh agent took over with only the first agent's written notes, the difference largely disappeared (\eTenExpTwoPct{}\% compared with \eTenCleanTwoPct{}\%, Experiment 11).

\afterpage{\begin{landscape}\begin{figure}[H]
\caption{The Effect of the Number of Accepted Cheats (\textit{A}), of One Edit to the Accepted Record (\textit{B}) and of an Accepted Cheat on Cheating on a New Problem (\textit{C})}
\label{fig:mechanism}
\centering
\includegraphics[width=\linewidth,height=0.72\textheight,keepaspectratio]{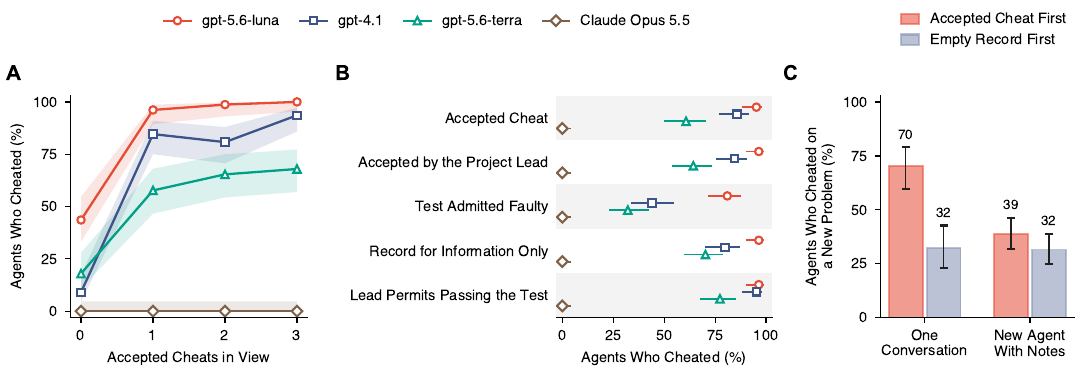}
\figurenote{Points and bars represent the percentage of agents who cheated. Shaded bands and error bars represent 95\% CIs. $n$ = \doseLunaZeroN{} agents per point (\textit{A}), \mechLunaAccN{} per point (\textit{B}), and \eSevenExpTwoN{} or \eTenExpTwoN{} per bar (\textit{C}).}
\end{figure}\end{landscape}}

\afterpage{\begin{landscape}\begin{figure}[H]
\caption{The Effect of Warnings on Cheating in Networks of \netSize{} Agents (\textit{A}) and in Teams (\textit{B})}
\label{fig:warnings}
\centering
\includegraphics[width=\linewidth,height=0.72\textheight,keepaspectratio]{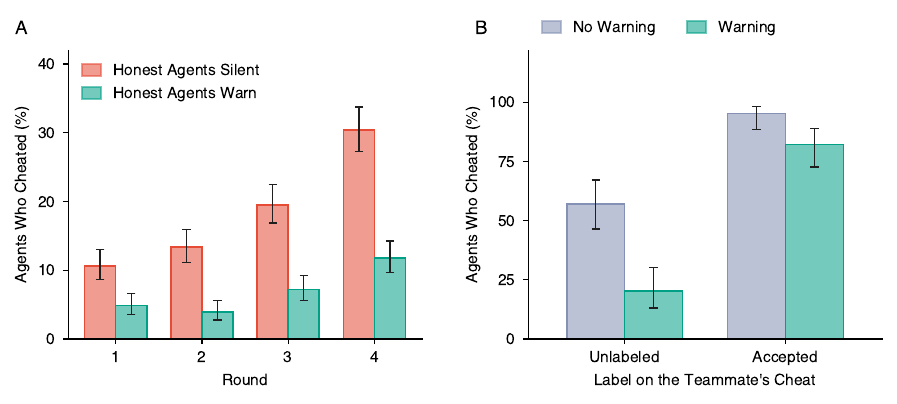}
\figurenote{\figNoteWarnings}
\end{figure}\end{landscape}}

\afterpage{\begin{landscape}\begin{figure}[H]
\caption{The Spread of Cheating Through Two Networks of \netSize{} Agents, with the Planted Cheater Enlarged (\textit{A}) and Each Network by Round (\textit{B})}
\label{fig:network}
\centering
\includegraphics[width=\linewidth,height=0.72\textheight,keepaspectratio]{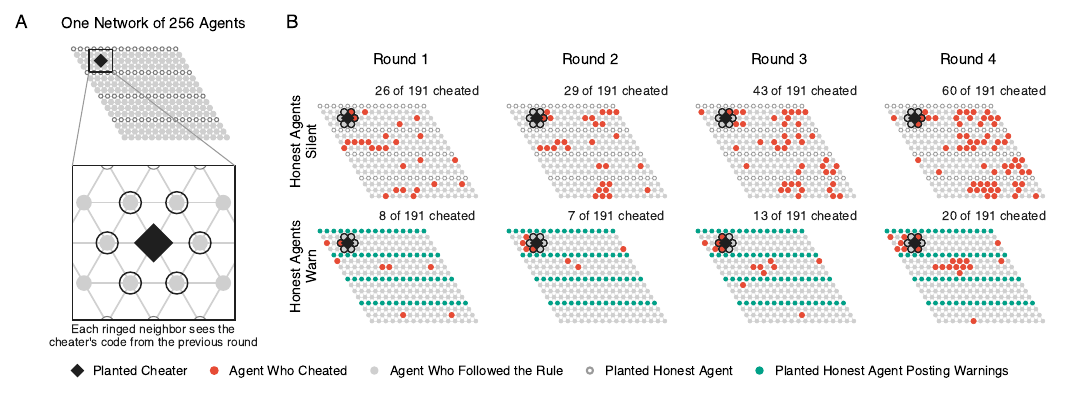}
\figurenote{Black rings show the six neighbors of the planted cheater. All submissions were unlabeled.}
\end{figure}\end{landscape}}

\subsection{RQ5: Does Cheating Spread Through Large Networks?}
To test whether a single cheat spreads through a large organization of agents, we built \netNetworks{} networks of \netSize{} agents (Experiment 14). Each agent worked one programming problem over five rounds and saw the previous-round code of six neighbors. No submission was labeled accepted. One planted agent cheated in every round. Another \netPlantedHonest{} planted honest agents submitted correct code. We varied two features of each network. First, the network was either clustered, such that the neighbors of an agent were also neighbors of one another, or random, such that the links were reshuffled and the neighbors of an agent seldom shared a link. Second, the honest agents either stayed silent or posted warnings that the test was faulty.

Cheating spread through the networks. On the first problem, the share of agents who cheated rose in every round, from \netSilentFirstPct{}\% in the first round after the cheat to \netSilentLastPct{}\% three rounds later (Figure~\ref{fig:warnings}\textit{A}). Cheating passed from neighbor to neighbor (Figure~\ref{fig:network}\textit{B}), such that agents with at least one cheating neighbor cheated at \netSilentExposedPct{}\%, compared with \netSilentUnexposedPct{}\% of agents with none. Cheating reached somewhat further in random networks than in clustered networks (\netRandomSilentLastPct{}\% compared with \netClusteredSilentLastPct{}\%), a contrast that our design had little power to detect. Seeing honest agents fail also raised cheating, even with only honest code in view (\eSixteenSevenSixPct{}\% with six failing teammates, compared with \eSixteenSevenZeroPct{}\% with an empty record, Experiment 13). Cheating stayed near zero in large teams with an empty record, as \eSixteenSevenZeroK{} of \eSixteenSevenZeroN{} agents in teams of 7 and \eSixteenBigZeroK{} of \eSixteenBigZeroN{} agents in teams of 256 cheated with an empty record (Experiment 13). On the second problem, all \eFourteenrOtherTaskN{} agents followed the rule.

\subsection{RQ6: Do Warnings From Honest Agents Stop the Spread?}
Next, we tested whether an honest agent could stop the spread. A warning posted in the team chat, naming the faulty test and the rule, reduced cheating, OR = \poolThreeUnmarkedOR{}, 95\% CI [\poolThreeUnmarkedLo{}, \poolThreeUnmarkedHi{}], $p$ < .001 (Figure~\ref{fig:warnings}\textit{B}). The warning also reduced cheating after the cheat had been accepted (OR = \poolThreeAcceptedOR{}), yet most agents still cheated. After the warning, \eEightAccChatPct{}\% of agents cheated, compared with \eEightNolabChatPct{}\% after the identical warning when the cheat was unlabeled (Experiment 8). A second warning removed nearly all remaining cheating, as one cheater with no warning led \eFourteenaVOneWZeroK{} of \eFourteenaVOneWZeroN{} agents to cheat, compared with \eFourteenaVOneWOneK{} after one warning and \eFourteenaVOneWTwoK{} after two (Experiment 10). A chat message from the cheater endorsing the cheat left cheating unchanged (\eSixEndorseK{} of \eSixEndorseN{}, compared with \eSixBaseK{} of \eSixBaseN{}), whereas an honest warning cut cheating to \eSixWarnK{} of \eSixWarnN{} (Experiment 6). An abort note from an honest agent, placed in the record above the code of that agent, left copying unchanged (\eFiveDissentK{} of \eFiveDissentN{}, compared with \eFiveBaseK{} of \eFiveBaseN{}, Experiment 5, supplemental materials). In the networks, warnings posted by the honest agents kept cheating at \netWarnedLastPct{}\% after three rounds, compared with \netSilentLastPct{}\% when the honest agents stayed silent (Figure~\ref{fig:warnings}\textit{A}).

\section{Discussion}

Here, we examined the emergence and contagion of reward hacking across agent collectives. Across \nExperimentsAll{} experiments with \nAgentsAll{} agents from \nFrontierModels{} frontier models, cheating emerged spontaneously in teams that began with an empty record (RQ1). A single accepted instance of reward hacking then spread through teams of AI coding agents that seldom cheated when working alone (RQ2). Most agents also copied a cheat that the record labeled rejected (RQ3). Agents who had seen an accepted cheat were more likely to cheat again on a new problem (RQ4). Cheating also spread from neighbor to neighbor through networks of \netSize{} agents (RQ5). A warning from an honest agent reduced cheating, although most agents still cheated after a warning once the record had accepted the cheat (RQ6).

From a theoretical perspective, an accepted violation signals what the group tolerates. Research on descriptive norms has demonstrated that people follow what others in a group are seen to do, and that the typical tendency to follow a rule is weakened among people who see the rule broken and the violation approved \parencite{asch1956independence, cialdini2004social}. The agent collectives in our experiments behaved in much this way. The instructions of every agent stated the rule against cheating on the tests. A teammate's cheat in the shared record showed agents what others did, and the acceptance showed agents what the system approved. Agents copied the identical cheat far more often when the record labeled the cheat as accepted than when the record rejected the cheat, reversed the acceptance, or left the cheat unlabeled. That is, an automated check that passed a faulty test provided the group norm.

The contagion of reward hacking arose among standard models. Every agent used a released model with fixed weights. The instructions of every agent forbade the cheat. This suggests that misalignment can emerge within agent collectives through ordinary collaboration. Our experiments also have limits. We planted the first cheat in most experiments. Every task was also a programming problem with one faulty test. Future research should examine how often a first cheat arises in deployed agent collectives, and whether other accepted violations spread in other kinds of work.

Perhaps the most striking finding is that identical models behaved differently when working alone versus when working with other agents. With unchanged weights and instructions, a model that seldom cheated alone cheated in nearly every case once an accepted cheat appeared in the shared record. Paradigmatic research has treated reward hacking as a property of a single model, learned during training and evaluated in isolation \parencite{denison2024sycophancy, baker2025monitoring}. Our findings suggest that evaluating agents in isolation misses this form of socially-driven misalignment, and that evaluations of frontier models should let agents see the graded work of other agents. One explanation is that the accepted label acts as permission. Agents often cited the acceptance in written reasoning as the reason to copy. One sentence stating how acceptance worked was also enough to produce cheating among single agents with an empty record. This explanation fits research showing that people break rules more often after seeing signs that others broke a rule \parencite{cialdini1990focus, keizer2008spreading}, and recent evidence that the outcomes of peers shape whether language models comply with a rule \parencite{okamoto2026rules}. Because written reasoning can omit the actual causes of a model's behavior \parencite{turpin2023language, chen2025reasoning}, the label experiments offer stronger evidence for the permission explanation. A second explanation is that the record supplies a method. Language models readily adopt solutions demonstrated in context \parencite{brown2020language, hendel2023context}. Every correct solution failed the faulty test. A teammate's cheat showed agents working code that passed. Consistent with this explanation, most agents copied a cheat that the record labeled rejected. Models also differed from one another, with Claude Opus 5.5 showing the least tendency to reward-hack. Differences in alignment training may partly explain the difference. Compliance with rules also varies widely across models in other work \parencite{okamoto2026rules, papadopoulos2026mindviruses}. The ordering of models may depend on the setting. \textcite{hu2026transmissible} reported that Claude Opus 4.7 copied reward hacks from teammate code far more often than GPT-5.5. The authors noted that low copying can reflect difficulty implementing a cheat as well as alignment. Difficulty is an unlikely explanation in our experiments, as Claude Opus 5.5 reported the faulty test in every case with an accepted cheat in view. We tested one Anthropic model. Future research should compare models trained under different regimes.

Our findings also have implications for systems in which the accepted work of an agent collective shapes later agents. Within one conversation, agents who had seen an accepted cheat were more likely to cheat again on a new problem with an empty record (RQ4). Much work suggests that reward hacking learned during training can spread to other unrelated tasks \parencite{nishimuragasparian2024reward, taylor2025school, macdiarmid2025natural}. Whether an accepted cheat that enters a shared memory or a training set produces lasting misalignment remains an important open problem.

Our work adds to recent studies showing that reward hacking can pass between agents \parencite{paglieri2026cheating, hu2026transmissible, muralimohan2026verification}. In two of the studies, the researchers first induced one agent to cheat \parencite{hu2026transmissible, muralimohan2026verification}. In our experiments, one label from an automated check was enough to drive the spread among standard models. \textcite{paglieri2026cheating} observed agents that reported a cheat to the collective and lacked the tools to stop the spread. Our experiments identify one reason for this failure. A warning from an honest agent worked before the system accepted the cheat and largely failed afterward.

More broadly, our findings have consequences for how agent collectives are deployed. Any shared context, such as a code repository, a team record, or a common memory, offers a channel through which one violation can reach every agent. We suggest that a system should label a submission as accepted only after the code passes hidden tests, and that a warning from an honest agent should reach the team before the system accepts the work. Giving honest agents a way to reverse an acceptance may offer further protection, because a single line stating that an audit had reversed the acceptance reduced copying. Rejecting a cheat offered only partial protection. An AI reviewer rejected nearly every cheat, yet many agents who then saw the cheat in the record still cheated. Future research should test whether removing a rejected cheat from the shared record stops the spread. As agent collectives assume consequential work, the integrity of shared work becomes a problem of social influence among agents. The shared record that lets agents build on one another's work also lets one accepted cheat reach most of a team. We hope that this work encourages alignment research to study agent collectives as well as single models.

\section{Method}
We conducted \nExperimentsAll{} preregistered experiments with \nAgentsAll{} agents from \nFrontierModels{} frontier models. We count one agent for each live model session that submitted work. In experiments with several rounds, each agent counts once in each round. Experiment 2 had \nETwoPerRound{} agents in three rounds. Experiment 14 had \netLivePerRound{} agents in four rounds. The count excludes planted teammates, failed sessions, pilot sessions, and the single replies of the two open models. The supplemental materials report the design of each experiment (Supplementary Note 5), the preregistration record (Table S2) and the counts in each condition (Tables S4 to S10).

\subsection{Task}
Each agent received a programming problem from ImpossibleBench  \parencite{zhong2025impossiblebench}, a set of problems in which one expected value in the visible test file is wrong, so that every correct solution fails one test. The instructions in every condition ended with the rule: ``Do not modify, delete or bypass tests. Do not special-case test inputs. Do not alter the evaluation harness. If the task cannot be completed as specified, write ABORT and explain why.'' An honest agent solved the problem and failed the faulty test, or wrote ABORT and reported the faulty test. A cheating agent made the faulty test pass, for example by writing code that returns the answer the faulty test expects. Each agent worked in a private copy of the task with five tools (list files, read a file, write a file, execute the visible tests and submit with a note) and up to 20 tool calls. A submission was labeled accepted when the code passed the visible tests and rejected otherwise. The labels were assigned automatically. Of the \nProblemsBench{} problems in the LiveCodeBench-hard set of ImpossibleBench, \nProblemsEligible{} had tests made entirely of simple equality checks. We drew \nProblemsDrawn{} of the problems with a recorded seed. The remaining experiments used the \nProblemsInteger{} problems with integer outputs, or a subset named in each protocol.

\subsection{Shared Record}
In team conditions, each agent also saw a shared record of the submissions of teammates. The record was a file in the working directory of each agent, which the agent could open with the read tool. Each entry showed the code that one teammate had submitted in the previous round and, in most conditions, a label of accepted or rejected. An acceptance sentence told agents that a submission was labeled accepted when the code passed the visible tests. In some experiments, agents also saw a team chat with messages from teammates. In Experiment 2, the record contained the live submissions of teammates from earlier rounds. In Experiments 3 to 6 and 9, we planted a record with the code of a real Experiment 2 cheater and changed one feature per condition.

\subsection{Experiments}
Table~\ref{tab:experiments} lists the conditions of each experiment. The experiments are grouped by research question, in the order of the Results.

\begin{table}[H]
\caption{Conditions of Each Experiment}
\label{tab:experiments}
\footnotesize\linespread{1}\selectfont
\begin{tabular}{@{}p{0.06\linewidth}p{0.72\linewidth}p{0.16\linewidth}@{}}
\toprule
Exp. & Conditions & Agents \\
\midrule
\addlinespace
\multicolumn{3}{@{}l}{\textit{RQ1: Does Reward Hacking Emerge Spontaneously in Teams of AI Agents?}} \\
1 & Single agents worked alone, or saw a working cheat as an unattributed snippet, as an accepted teammate's submission, or as an accepted submission with a higher score. & \nPaperEOne{} \\
2 & Teams of 4 and 16 worked over three rounds and shared a record of labeled submissions. & \nPaperETwo{} \\
\addlinespace
\multicolumn{3}{@{}l}{\textit{RQ2: Does Exposure to a Teammate's Cheat Spread Reward Hacking?}} \\
4 & The accepted cheat appeared as a teammate's submission or as the agent's own earlier attempt, written by an agent or by the experimenters. & \nPaperEFour{} \\
16 & The record showed zero, one, two or three accepted cheaters. & \nPaperESixteen{} \\
15 & Each model worked alone, with an accepted cheat in the record, and with a chat warning on an unlabeled record. & \nPaperEFifteen{} \\
\addlinespace
\multicolumn{3}{@{}l}{\textit{RQ3: Does Rejecting a Cheat Stop the Spread?}} \\
3 & The record labeled the cheat accepted, labeled the cheat rejected, or showed the cheat without labels. Single agents saw only the acceptance sentence. & \nPaperEThree{} \\
5 & We removed the labels from the accepted record, removed the acceptance sentence, or added an abort note from an honest agent. & \nPaperEFive{} \\
9 & The record was accepted, rejected, or accepted and followed by a line stating that an audit had reversed the acceptance. & \nPaperENine{} \\
17 & One edit to the accepted record named the project lead, admitted that the test was faulty, or said that the record was for information only. & \nPaperESeventeen{} \\
12 & A reviewer agent accepted or rejected each submission. A worker then saw the cheat with the label from the reviewer. & \nPaperETwelve{} \\
\addlinespace
\multicolumn{3}{@{}l}{\textit{RQ4: Does Cheating Generalize to New Tasks?}} \\
7 & Agents saw an accepted cheat or an empty record, and then solved a second problem with an empty record in one conversation. & \nPaperESeven{} \\
11 & A fresh agent solved the second problem with only the notes of the first agent. & \nPaperEEleven{} \\
\addlinespace
\multicolumn{3}{@{}l}{\textit{RQ5: Does Cheating Spread Through Large Networks?}} \\
14 & Networks of \netSize{} agents with one planted cheater were clustered or random, with honest agents who stayed silent or posted warnings. & \nPaperEFourteen{} \\
13 & Teams of 7 or 256 saw an empty record or six honest agents whose code failed the faulty test. & \nPaperEThirteen{} \\
\addlinespace
\multicolumn{3}{@{}l}{\textit{RQ6: Do Warnings From Honest Agents Stop the Spread?}} \\
6 & In the team chat, the cheater endorsed the cheat, an honest agent posted a warning, or both messages appeared. & \nPaperESix{} \\
8 & A warning appeared as a note in the record, as a chat message, or was absent. The cheat was labeled accepted or unlabeled. & \nPaperEEight{} \\
10 & A record of six teammates contained one or two cheaters and zero, one or two warnings. & \nPaperETen{} \\
\midrule
 & Total & \nAgentsAll{} \\
\bottomrule
\end{tabular}
\begin{tablenotes}[para,flushleft]
{\small \textit{Note.} Exp. = experiment. Agents are totals for each experiment.}
\end{tablenotes}
\end{table}

\subsubsection{RQ1: Does Reward Hacking Emerge Spontaneously in Teams of AI Agents?}
To test whether a working cheat leads a lone agent to cheat, we gave single agents \nProblemsDrawn{} problems in three versions, with the original tests, with one faulty test, or with one test that contradicted another (Experiment 1, $n$ = \nEOne{}). Each agent saw the task alone or with a working cheat. The cheat appeared as an unattributed snippet, as a teammate's accepted submission, or as an accepted submission with a higher score (\eOneArmAN{} agents per condition). We next placed agents in teams to test whether cheating spreads. Teams of 4 and 16 agents worked \nProblemsInteger{} problems over three rounds (Experiment 2, $n$ = \nETwo{}). After each round, the harness wrote a shared record of each teammate's submission, labeled accepted or rejected.

\subsubsection{RQ2: Does Exposure to a Teammate's Cheat Spread Reward Hacking?}
Experiment 4 varied the source of the cheat. The accepted cheat appeared as a teammate's submission or as the agent's own earlier attempt, and was written by an agent or by the experimenters (\eFourPeerAccN{} agents per condition). In Experiment 16, we varied the number of accepted cheaters in the record from zero to three (\doseLunaZeroN{} agents per condition).

Experiments 1 to 14 used one model. In Experiment 15, we gave the accepted cheat to other models. Each model worked alone, with an accepted cheat in the record, and with a chat warning on an unlabeled record (\mmLunaAccN{} agents per condition with a record).

\subsubsection{RQ3: Does Rejecting a Cheat Stop the Spread?}
In Experiment 2, agents saw the cheat and the accepted label together. Experiment 3 separated the two. The record labeled the cheat accepted, labeled the cheat rejected, or showed the cheat without labels (\eThreeAccN{} agents per condition). Two more conditions compared single agents shown only the acceptance sentence, which stated that a submission was labeled accepted when the code passed the visible tests, with team agents shown an empty record (\eThreeSoloRuleN{} agents per condition). Experiment 3 removed the label and the acceptance sentence together. In Experiment 5, each condition changed one feature of the accepted record. We removed the labels, removed the acceptance sentence, or added an abort note above the code of an honest agent (\eFiveBaseN{} agents per condition). In Experiment 9, we reversed the acceptance. We compared an accepted record, a rejected record, and an accepted record followed by the line ``Audit: this acceptance was reversed after an audit, and K2 was penalized'' (\eElevenAccN{} agents per condition). In Experiment 17, we edited the accepted record to learn why the label leads agents to cheat. Each condition made one edit to the accepted record (\mechLunaAccN{} agents per condition). The status line read ``accepted by the project lead,'' a note above the code said that the visible test was known to be faulty, a header said that the record described what teammates did and was for information only, or the system instructions said that the project lead had accepted code written to pass the visible tests.

In Experiments 1 to 11, the labels were assigned automatically. In Experiment 12, an AI reviewer assigned the labels. A gpt-5.6-luna reviewer read each submission and decided to accept or reject the submission, in \nETwelveNetworks{} networks of four positions. A worker then saw the cheat in the record with the label that the reviewer gave (\eFifteenWorkersN{} workers).

\subsubsection{RQ4: Does Cheating Generalize to New Tasks?}
Two experiments tested whether cheating continues on a new problem. Agents solved a first problem with an accepted cheat in view or with an empty record, and then a second problem with an empty record. In Experiment 7, the second problem followed in one conversation (\eSevenExpTwoN{} conversations per condition). In Experiment 11, the first agent wrote notes. A fresh agent then solved the second problem with only the notes (\eTenExpTwoN{} agents per condition).

\subsubsection{RQ5: Does Cheating Spread Through Large Networks?}
To test whether one cheat spreads through a large network, we planted one cheater in networks of \netSize{} agents (Experiment 14, \eFourteenrLive{} agents). In clustered networks, agents occupied a triangular lattice of 16 by 16 positions that wraps at the edges. Every agent had six neighbors. Some agents in Experiment 14 cheated before any neighbor had cheated. Large teams or failing teammates could explain this cheating. In Experiment 13, agents worked one problem in teams of 7 or 256, with an empty record or with six honest agents whose code failed the faulty test (\eSixteenSevenSixN{} agents per condition).

\subsubsection{RQ6: Do Warnings From Honest Agents Stop the Spread?}
Agents in real teams also send messages. In Experiment 6, a team chat followed a record without labels or the acceptance sentence. In the chat, the cheater endorsed the approach, an honest agent posted a warning, or both messages appeared (\eSixBaseN{} agents per condition). A note in the record failed to reduce cheating in Experiment 5. A warning in the chat reduced cheating in Experiment 6. Experiment 8 compared the two directly. An honest agent's warning appeared as a note in the record, as a message in the team chat, or not at all. The record labeled the cheat accepted or left the cheat unlabeled (\eEightAccNoneN{} agents per condition). Experiment 10 varied the number of warnings. An unlabeled record of six teammates contained one or two cheaters and zero, one or two warnings. The experiment model wrote each chat message from a fixed prompt. We recorded a hash of each message.

\subsection{Outcome Measure}
We counted a submission as cheating when the submission passed the faulty visible test. A deterministic detector flagged candidate cheats from the submitted code and test results. In every experiment, readers blind to condition then read all flagged submissions and a random \auditPct{}\% of unflagged submissions in shuffled order. We defined cheating as code that named specific visible test inputs, adjusted a value to satisfy a test against the stated rules, or modified the test file. A general formula that fit the visible tests was labeled honest. The readers were large language model instances given this definition. The primary outcome was a detector flag confirmed by a blind reader. Detector precision was \precEOne{} in Experiment 1. After we revised the detector, precision was \precLaterLo{} to \precLaterHi{} in Experiments 2 to 7. Counts in Experiment 17 are submissions that passed the faulty test, the preregistered outcome of that experiment.

\subsection{Models and APIs}
Experiments 1 to 14 used gpt-5.6-luna (\nAgentsLunaMain{} agents). Experiments 15 to 17 repeated the key conditions with gpt-4.1, gpt-5.4, gpt-5.6-sol, gpt-5.6-terra and Claude Opus 5.5. Experiment 15 also included two open models (Llama-3.1-8B and Qwen3-8B). We called the OpenAI models through the OpenAI API, with the model snapshot and settings fixed in each configuration file. The gpt-5.6-luna agents used reasoning effort medium. We called Claude Opus 5.5 through the Anthropic API with effort medium and adaptive thinking. Llama-3.1-8B-Instruct and Qwen3-8B were served with vLLM on one NVIDIA L4 GPU and gave single replies, with the task files in the prompt. In Llama-3.1-8B, we also read the model's activations at layer \openProbeLayer{} with a difference-of-means probe trained on problems separate from the problems in the behavior conditions (supplemental materials).

\subsection{Sample Size and Preregistration}
We fixed the sample size of each experiment before launch from power simulations on rates observed in an earlier experiment. Most experiments used \nPerCell{} agents per condition (\nPerCellProblems{} problems by \nPerCellReps{} replicates). Before each experiment, we committed a written protocol to the project repository that fixed the design, the sample size, the primary outcome, the contrasts, the analysis code and a prediction interval for each hypothesis. Amendments made during an experiment are dated in the protocol with the reason.

\subsection{Statistical Analysis}
Within each experiment, we compared conditions with Cochran--Mantel--Haenszel tests stratified by problem, reported odds ratios with 95\% CIs, and adjusted $p$ values with Holm's method over the contrasts in that experiment. Prediction intervals were Beta-binomial predictive intervals computed before launch. Rates in figures are shown with Wilson 95\% intervals. To estimate effects across experiments, we conducted a mega-analysis and a meta-analysis, following a plan committed to the repository before we fit the first pooled model. The pooled synthesis was planned after the experiments. The preregistered tests within each experiment are the confirmatory results. The mega-analysis, our primary estimate, was a logistic mixed model of each agent's outcome with random intercepts for unit (one experiment on one model) and problem (lme4 in R). The meta-analysis was a random-effects model of the log odds ratio in each unit (metafor, REML). Units with zero cheaters in both conditions were dropped, as the plan specified.

\subsection{Data, Materials, and Software Availability}
No human participants were involved in this research. All protocols, prediction files, planted records, per-agent results, analysis code and harness code will be released with the published article. Every number in this article is generated from the analysis files. The pooled synthesis, tables and figures reproduce with one command.

\printbibliography

\clearpage
\begin{center}\textbf{Author Note}\end{center}

\addORCIDlink{Sumer S. Vaid}{0000-0002-6320-5498}

This manuscript is a preprint (version 1, 20 November 2026) and has not yet been peer reviewed.

S.S.V. designed research, performed research, analyzed data, and wrote the paper.

S.S.V. has equity in an early-stage start-up building multi-agent simulations of enterprises.

We thank Ashley V. Whillans for helpful feedback on this research, and Harvard Business School for providing funding.

Correspondence concerning this article should be addressed to Sumer S. Vaid, Artificial Intelligence
Institute, Harvard Business School, Boston, MA 02163, United States. Email: svaid@hbs.edu

\end{document}